\documentclass[conference]{IEEEtran}

\IEEEoverridecommandlockouts
\usepackage{cite}
\usepackage{amsmath,amssymb,amsfonts}
\usepackage{algorithm}
\usepackage{algpseudocode}
\usepackage{graphicx}
\usepackage{textcomp}
\usepackage{blindtext}
\usepackage{subfigure}
\usepackage{float} 
\usepackage{subfigure}
\usepackage{xcolor}
\usepackage{diagbox}
\usepackage{slashbox}
\usepackage{wrapfig}
\usepackage{makecell}
\def\BibTeX{{\rm B\kern-.05em{\sc i\kern-.025em b}\kern-.08em
		T\kern-.1667em\lower.7ex\hbox{E}\kern-.125emX}}
\makeatletter

\begin{document}
	\title{Mitigating Unnecessary Handovers in Ultra-Dense Networks through Machine Learning-based Mobility Prediction}
	\IEEEpeerreviewmaketitle
	\author{\IEEEauthorblockN{Donglin Wang\IEEEauthorrefmark{2}, Anjie Qiu\IEEEauthorrefmark{2}, Sanket Partani\IEEEauthorrefmark{2}, Qiuheng Zhou\IEEEauthorrefmark{1},  and Hans D. Schotten\IEEEauthorrefmark{2}\IEEEauthorrefmark{1}}
		\IEEEauthorblockA{\textit{\IEEEauthorrefmark{2}Rhineland-Palatinate Technical University of Kaiserslautern-Landau, Germany} \\
		$\{$dwang,qiu,partani,schotten$\}$@eit.uni-kl.de \\} 
		\IEEEauthorblockA{\textit{\IEEEauthorrefmark{1}German Research Center for Artificial Intelligence (DFKI GmbH), Kaiserslautern, Germany} \\
		$\{$qiuheng.zhou,schotten$\}$@dfki.de}
	}
	\maketitle

\begin{abstract}
In 5G wireless communication, Intelligent Transportation Systems (ITS) and automobile applications, such as autonomous driving, are widely examined. These applications have strict requirements and often require high Quality of Service (QoS). In an urban setting, Ultra-Dense Networks (UDNs) have the potential to not only provide optimal QoS but also increase system capacity and frequency reuse. However, the current architecture of 5G UDN of dense Small Cell Nodes (SCNs) deployment prompts increased delay, handover times, and handover failures. In this paper, we propose a Machine Learning (ML) supported Mobility Prediction (MP) strategy to predict future Vehicle User Equipment (VUE) mobility and handover locations. The primary aim of the proposed methodology is to minimize Unnecessary Handover (UHO) while ensuring VUEs take full advantage of the deployed UDN.  We evaluate and validate our approach on a downlink system-level simulator. We predict mobility using Support Vector Machine (SVM), Decision Tree Classifier (DTC), and Random Forest Classifier (RFC). The simulation results show an average reduction of 30\% in handover times by utilizing ML-based MP, with RFC showing the most reduction up to 70\% in some cases.  

\end{abstract}


\section{introduction}
The fifth generation (5G) of cellular communication technology aims to achieve a significant enhancement in communication capacity \cite{alsharif5g2017}. One of the key innovations in 5G is the implementation of UDN, which involves the deployment of a high density of small cells in order to support the efficient utilization of the spectrum and augment the coverage and capacity of cellular networks. Therefore, wide attention has been paid to the application of UDNs. VUEs have a collection of flexible mobile access options provided by UDN with higher spectrum efficiency through the extensive deployment of SCNs \cite{udnsurveylopez2015}. 
\begin{figure}[htbp]
	\centering
	\includegraphics[width=\linewidth]{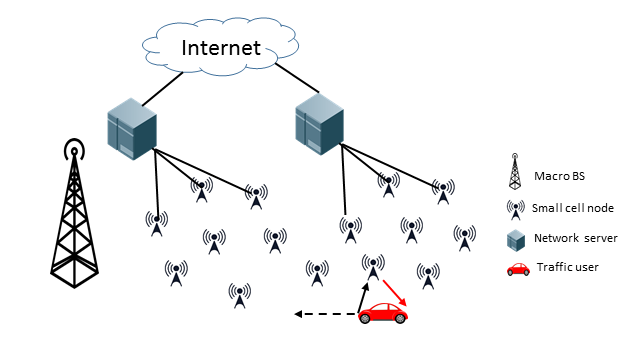}
	\caption{A showcase of a UDN network}
	\label{fig}
\end{figure}

The UDN network consists of a large number of small cells that can support spatial reuse of the spectrum to increase the coverage and capacity of cellular networks remarkably \cite{udnsurveylopez2015}. As shown in Fig. 1, a simplified UDN network has multiple SCNs, a macro Base Station (BS), a network server, a network, and a moving VUE. Densely deployed SCNs improve the frequency reuse and system capacity, and more VUEs and mobile users can be supported by UDN\cite{tttdensitywang2022}. Nonetheless, the deployment of small cells differs from that of traditional BSs, and the mobility in UDN poses unique challenges in comparison to previous multi-tier networks. UDNs are faced with a plethora of obstacles, including increased interference, more frequent handovers, delays in traffic delivery, radio link failures, and elevated costs \cite{surveysotayu2016}.
Literature has been devoted to providing insights on the challenges associated with UDN deployment and identifying potential solutions \cite{gotsisudn2016}\cite{udntrendshao2016}. In this work\cite{chenudn2018}, the UDN security problem is considered. The authors proposed an Implicit Certificate (IC) scheme that is expected to solve the UDN security problem among the Access Point (AP) in a dynamic APs group and between the AP and User Equipment (UE).

In \cite{3gpp36331}, the Third Generation Partner Program (3GPP) has defined the handover processes protocol. Handover events are generally triggered based on periodic measurement reports of VUEs and handover parameters, i.e., Time-To-Trigger (TTT) and Handover Hysteresis value (Hys) \cite{tttdensitywang2022}. In our previous work \cite{tttdensitywang2022}, the authors analyze the effect of TTT, density, and velocity on the performance of handovers. In \cite{pphoalraih2021}, the authors proposed a handover optimization method that uses fuzzy logic to reduce handover probability. The simulation results show handover probability and handover Ping-Pong (PP) effects have been significantly reduced.  In \cite{raoshimobilityudn2019}, Shi et al. proposed a user MP method based on the Lagrange Interpolation. By using the slope of the trajectory polynomial and velocity, the transition probability of VUEs is detected. However, this method suffers from low mobility prediction accuracy and less UHO reduction. 
In \cite{selfadaptudnhuang2022}, the authors tried to minimize the handover failures and PP rate by introducing an intelligent dynamic handover parameter optimization approach.
With the development, more and more ML algorithms are used in 5G communication \cite{mo5g2019}. We propose ML algorithm-based MP to reduce the UHO times. Different algorithms are used and compared the prediction results. Then the predicted results are applied to reduce handover times. 

The remaining sections of this paper are organized as follows: Section II delves into the implementation of ML-based MP and provides an overview of the system model, the ML models employed, and the results obtained. Section III discusses the handover process in UDNs with and without MP in depth. Section IV introduces the system simulator and presents a comprehensive analysis of the results obtained. Finally, in Section V, the study concludes with a summary of the findings and suggestions for future research. 

\section{Machine learning-based mobility prediction}

With the advancement of modern society, mobile communication is increasingly required to operate in highly dynamic, pervasive computing environments on the road network. This demands increased capacity and exceptional QoS. To meet these requirements, ML-based MP has emerged as a critical enabler in mobile communication. MP utilizes historical traffic information to predict the future locations of VUEs. thereby enabling efficient radio resource management, route planning, vehicle dispatching, and reducing traffic congestion and handovers \cite{wangmobility2021}. In the following subsections, a detailed examination of  ML-based MP will be provided, including the system model and various ML algorithms employed.

\begin{figure}[htbp]
	\centering
	\includegraphics[width=\linewidth]{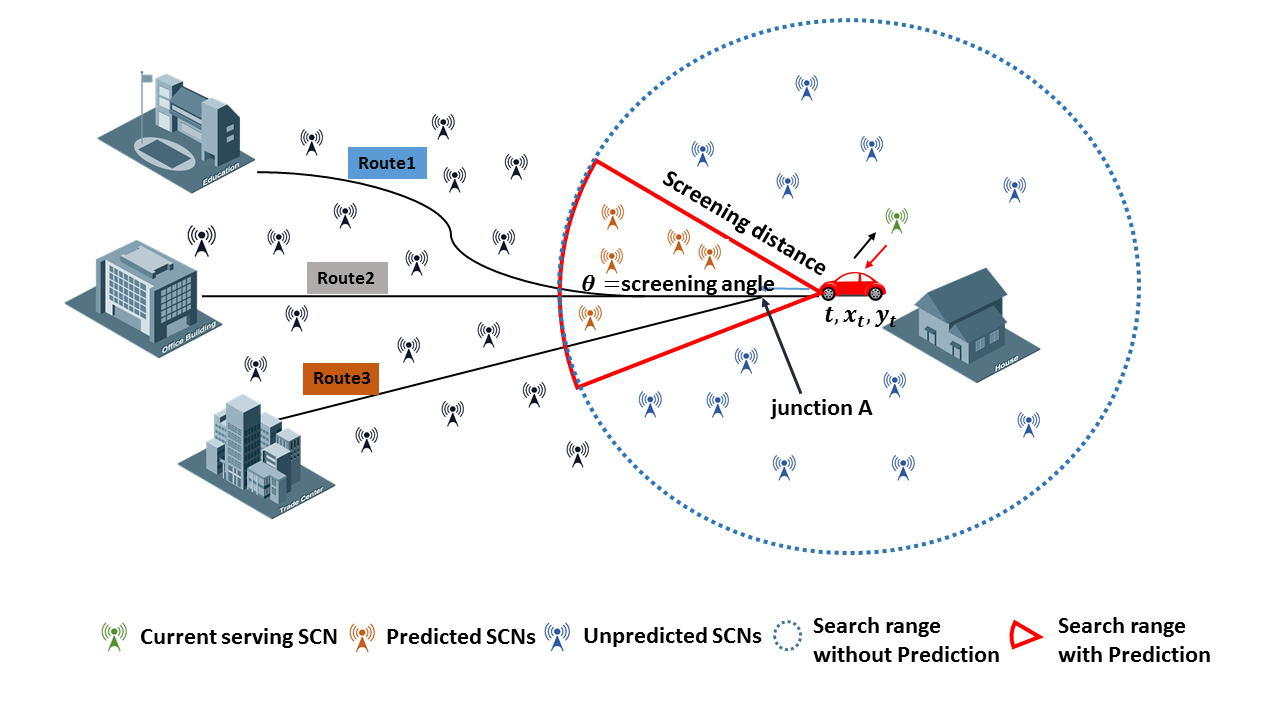}
	\caption{VUE searching range with and without prediction}
	\label{VUE with and without MP}
\end{figure}

\subsection{System models}
As depicted in Fig. \ref{VUE with and without MP}, a simulation scenario of UDN with SCNs is constructed. The deployment of SCNs in UDN adheres to the Poisson Point Process (PPP) distribution model \cite{lastppp2017}. It's assumed that all SCNs possess identical characteristics, including antenna height, antenna tilt, and others. 

\begin{figure}[htbp]
	\centering
	\includegraphics[width=\linewidth]{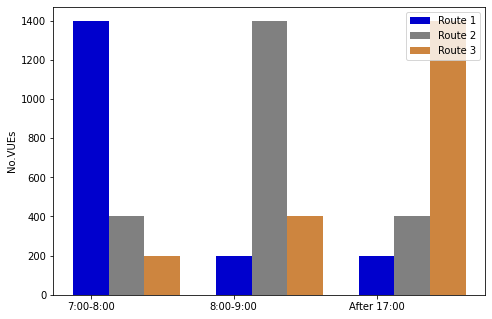}
	\caption{User distribution}
	\label{user_dis}
\end{figure}

In this system model, three distinct routes are available for thousands of VUEs within a 1000-meter *1000-meter area \cite{nandish2020handover}. 
Fig. \ref{user_dis} illustrates the distribution of VUEs on potential routes, varying mobility demands depending on the time of the day. 

For example, Route 1 is frequently utilized by VUEs during the morning time commute from 7-8 AM for transportation to school. Conversely, VUEs may opt for Route 2 between 8-9 AM to reach workplaces. For instance, between 7-8 AM, it's simulated that 1400 VUEs take Route 1 and 400 VUEs select Route 2, and the remaining 200 VUEs take Route 3. Similar VUE distributions are observed over the other time periods. The simulation scenario and moving VUEs are generated using SUMO simulation software \cite{sumointro2011}. VUEs are classified after passing junction A, i.e., the VUEs on Route 1 are marked with the ID ”0”. VUEs on Route 2 are identified with the ID ”1”, and  the VUEs on Route 3 is labeled with the ID "2". 

\subsection{Dataset for training}
Each vehicle is simulated for 70000 milliseconds, and the contextual information of each vehicle, such as time step, route ID, longitude and latitude of VUEs, is collected. The generated dataset $D=(x_i,y_i,r_i,t_i)$ is utilized to train various ML algorithms, where $x_i$ represents longitude, $y_i$ represents latitude, $r_i$ represents the route ID, and $t_i$ represents the time step. The dataset $D$ is randomly partitioned into a training dataset (75\%) and a testing dataset (25\%). The size of our dataset is $N=82435$ obtained from the SUMO simulation software.

\subsection{Machine learning algorithms}
\begin{figure}[htbp]
	\centering
	\includegraphics[width=\linewidth]{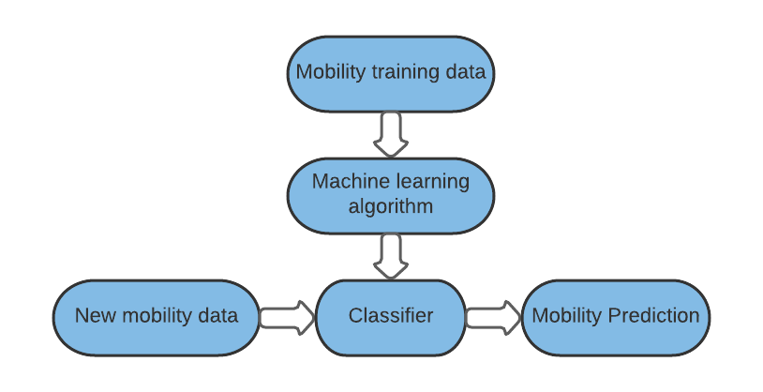}
	\caption{ML model for MP \cite{nandish2020handover}}
	\label{MP}
\end{figure}

The process of MP is illustrated in Fig. \ref{MP}. In this process, the collected dataset $D$ is used to identify the relationship between input data and the desired output value. When new input data $(x_k,y_k,t_k)$ is input into the model, the model can predict the next output value of route $(r_k)$ that $VUE_k$ will take at the time slot $t_k$. 
As the MP problem is a classification problem, a variety of ML models such as SVM, DTC, and RFC are employed for MP and compared in terms of performance for large-scale data training. 
\subsubsection{Support Vector Machine (SVM)}
SVMs are a widely used ML algorithm in classification and regression. As stated in  \cite{svmintro2006}, they have been widely adopted in the field of MP of VUE due to their ability to achieve high prediction accuracy while minimizing the risk of overfitting the data. In this study, we utilized the SVM algorithm to predict the VUE's route as previously demonstrated in \cite{wangmobility2021}.

\subsubsection{Decision Tree Classifier (DTC)}
DTCs are a popular ML algorithm that aims to simplify the complex decision-making process by breaking them down into a series of simpler decisions, hoping the final solution obtained this way would resemble the intended desired solution \cite{safaviandecisiontree1991}.
\subsubsection{Random Forest Classifier (RFC)}
RFCs are an ensemble technique that combines multiple decision trees to improve prediction accuracy. According to \cite{breimanrandomforests}, the technique works by having each decision tree vote for the most popular class, and the class that receives the majority of votes is chosen as the final output. This approach has been shown to result in significant improvements in classification accuracy when compared to using a single decision tree. 

\subsection{Performance comparison of ML algorithms}
In Table \ref{ML models comparison}, the performance comparison of three ML models is illustrated, in terms of five performance metrics: Training Set Score (TSS), Testing Set Score (TESS), Balanced Accuracy Score (BAS), Recall Score (RS), and Precision Score (PS). The TSS and TESS values indicate the model's performance on the training and testing sets, respectively. A comparable TSS and TESS suggest that there is no significant overfitting or underfitting. The BAS metric is used to quantify the model's overall performance by taking into account the recall obtained on each class, while the RS and PS metrics measure the model's ability to identify all positive samples, and avoid falsely labeling negative samples as positive respectively. SVM, DTC, and RFC used in this system all have a very high TSS with 0.9112, 1.0, and 1.0 respectively, indicating that the models have been trained very well on the data, and TESS values are also high, indicating that the models are able to generalize well to new data. Especially, RFC with the highest TESS value of 0.9606.

The trained ML models are used to predict the user route based on the historical data, for various time periods. Table \ref{prediction accuracy comparison} shows the TESS values (denoted as prediction accuracy) for the three different ML models. As an example, during the morning time period of 7-8 AM, the probability of a VUE taking Route 1 is 0.9113 when using the SVM model as opposed to 0.9063 when using the RFC model. The prediction results are used in the following sections for reducing handover.

\begin{table}[htbp]
\caption{ML models comparison for classification}
\begin{center}
\begin{tabular}{llllll} \hline
{\textbf{Model}}& {\textbf{TSS}}&{\textbf{TESS}}& {\textbf{BAS}}& {\textbf{RS}}& {\textbf{PS}}       \\ \hline
SVM   &0.9112  &0.9443  &0.7135 &0.9113 &0.9202  \\ 
DTC   &1.0     &0.9571  &0.8364  &0.8994 &0.8987  \\ 
RFC   &1.0     &0.9606  &0.7749  &0.9063 &0.9034  \\ \hline
\end{tabular}
\label{ML models comparison}
\end{center}
\end{table}

\begin{table}[htbp]
\caption{prediction accuracy comparison for different ML algorithms}
\begin{center}
\begin{tabular}{ |p{3.4cm}|p{0.7cm}|p{0.7cm}|p{0.7cm}|p{0.7cm}|  }
\hline
\backslashbox[38mm]{\textbf{Time Period}}{\textbf{P\_accuracy}} {\textbf{ML\_models}}
&\makebox{\textbf{Route}}&\makebox{\textbf{SVM}}&\makebox{\textbf{DTC}}&\makebox{\textbf{RFC}} \\
\hline
7:00-8:00   & Route1 & 0.9113  & 0.8994 & 0.9063 \\ \hline
8:00-9:00   & Route2 & 0.9580  & 0.9878 & 0.9916 \\ \hline
17:00-18:00 & Route3 & 0.9639  & 0.9840 & 0.9840 \\ \hline
\end{tabular}
\label{prediction accuracy comparison}
\end{center}
\end{table}

\section{Handover process with MP applied ML models}
The integration of SCNs in UDN has led to a proliferation of UHO. To mitigate this issue, we propose a new method that utilizes MP to reduce handover times. For comparison, the conventional 5G handover method is introduced as well. Simulation results of two methods are used to compare the performance on the handover. 

\subsection{Conventional 5G handover calculation} 

The conventional 5G handover calculation method is well-established, in which a user explores the Signal-to-Interference-plus-Noise Ratio (SINR) and availability of all nearby SCNs to discover or search for potential SCNs to connect to \cite{tayyabnrhandover2019}\cite{peltonen5ghandover2021}, as depicted in Fig. \ref{VUE with and without MP}. In this figure, the blue circle represents the communication range ($R$) between VUE and SCNs. If the distance ($D_{ki}$) between the $VUE_k$ and the $i_{th}$ SCN is less than or equal to the communication range, then the $i_{th}$ SCN will be considered as a candidate SCN. In total, $T$ candidates are available including the SCNs marked in orange and blue in Fig. \ref{VUE with and without MP}. The SCN that provides the highest SINR value will become the target SCN, and the VUE will be scheduled to connect to it, even though it may not be necessary. 
However, the highly dense deployment of SCNs in UDN leads to more UHO, thus increasing traffic delay, interference, radio link failure, and ultimately decreasing the user experience quality. 

\subsection{5G handover calculation with Prediction}
The proposed method for 5G handover optimization incorporates the utilization of predictive techniques to anticipate handover events. The method involves the implementation of a screening distance and screening angle, calculated as $D$ = 300 meters and $\theta = {50/\sqrt{V} + 18 }$ respectively, where $V$ represents the velocity of the VUE \cite{raoshimobilityudn2019}. The SCNs within the defined screening distance and angle are considered as predicted candidates for handover (denoted as $N$) in orange, with the remainder of the SCNs within the communication range referred to as unpredicted candidates (denoted as $M$). 

As illustrated in Fig. \ref{VUE with and without MP}, if a $VUE_k$ has the highest SINR value (denoted as $ki\_dBm$) from an unpredicted candidate ($M_i$) at a future position, the unpredicted candidate is labeled as the incorrect SCN. The highest SINR (denoted as $kj\_dBm$) value from the predicted candidate ($N_j$) is then compared to the SINR value (denoted as $kx\_dBm$) of the $VUE_k$ from the current serving SCN ($x$). An SINR offset value (denoted as $so\_dBm$), calculated as $so\_dBm = (ki\_dBm-max(kj\_dBm, kx\_dBm))*3$ is employed. Then the highest SINR ($ki\_dBm$) from the incorrect SCN becomes $ki\_dBm$ = $ki\_dBm$ - $so\_dBm$ to ensure that the $VUE_k$ doesn't switch unpredicted candidate ($M_i$) but switches to the predicted SCN ($N_j$) with the highest SINR $kj\_dBm$, thus optimizing the user experience and reducing handover times.

The proposed method for 5G handover triggering as 3GPP standardization is outlined in Algorithm I.

\begin{algorithm}
    \caption{Handover triggering logical algorithm}\label{euclid}
    \hspace*{\algorithmicindent} \textbf{Input}: serving\_scn, serving\_sinr, best\_sinr, target\_scn, sinr\_min, avg\_sinr, best\_cio, current\_cio, ho\_hys, ttt, ho\_timer, ho\_trigger, ho\_exec\_time    \\
    \hspace*{\algorithmicindent} \textbf{Output}: ho\_times
    \begin{algorithmic}[1]
    \If {$serving\_{scn} \neq target\_{scn}$} 
        \If{$best\_sinr > sinr\_min  \; \& \; best\_sinr - avg\_sinr + best\_cio - current\_cio > ho\_hys$}
         \State  $ho\_trigger\gets  1$
         \State  $ho\_timer  \gets  ho\_timer +1$
            \If{$ho\_timer == ttt$}
             \State $serving\_scn \gets  target\_scn$
             \State $ho\_exec\_time\gets  25 $
             \State $ho\_times \gets ho\_times + 1 $
             \State $ho\_trigger \gets  0$
             \State $ho\_timer   \gets  0$
            \EndIf
         \EndIf
    \EndIf
    \end{algorithmic}
\end{algorithm}
All simulation parameters are defined in Table \ref{simulation parameters}. As demonstrated in Algorithm I, the total number of successful handovers in each simulation is extracted. The handover triggering process is executed in the following steps:
1) It is essential to ensure that the $target\_scn$ is distinct from the current $serving\_scn$, meaning that they possess different identifiers and are located in disparate locations. This step ensures that the handover process is triggered only when a distinct target cell is identified, rather than the current serving cell, thus avoiding UHOs and ensuring optimal network performance; 2) This step involves evaluating the value of $best\_sinr$, which represents the best SINR value calculated previously, against a pre-defined threshold of $sinr\_min = -7 \, dB$. Additionally, the difference between $best\_sinr$ and $avg\_sinr$ pluses the difference of $best\_cio$ and $current\_cio$ must be greater than the handover hysteresis value of $ho\_hys = 3 \, dB$. The values of $best\_cio$ and $current\_cio$ are initially set to 0, however, they may be affected when a load balancing algorithm is implemented. $avg\_sinr$ represents the average SINR of the VUE in relation to the current SCN and is calculated using the previous $X$ SINR values, where $X$ represents the number of previous SINR values used, and taking their average. In this simulation, a counter is utilized to hold 10 previous SINR values and $avg\_sinr$ is calculated at each tic in the simulator. It can only be calculated when the previous X SINR values are available; 3) and 4) If all the conditions mentioned in the previous steps are satisfied, this step sets the $ho\_trigger$ flag to one, indicating that a handover should be triggered. Additionally, the $ho\_timer$ counter is incremented by one; 5) This step involves evaluating whether the value of the $ho\_timer$ counter has reached the predefined TTT value. If the TTT value has been met, it indicates that the handover should be executed. If the TTT value has not been met, the handover process will be delayed until the TTT value is reached. The TTT value is a crucial parameter that affects the handover rate, as it determines the time interval between the initiation of the handover process and its execution; 6) This step marks the execution of the handover process. The $serving\_scn$ is updated to equal the $target\_scn$, indicating that the handover has taken place. The $ho\_exec\_time$ is set to 25 tics, which represents the duration of the handover process. The output counter $ho\_times$ is incremented by 1 to keep track of the number of successful handovers. The $ho\_trigger$ and $ho\_timer$ are reset to 0 to prepare for the next handover process. In case one of the conditions mentioned in the previous steps is not satisfied, the algorithm will be re-executed. The above description outlines the basic logic for the handover triggering process in each simulation.

\begin{table}[htbp]
\caption{Simulation parameters}
\begin{center}
\begin{tabular}{ |p{3cm}|p{5cm}|  }
 \hline
    \textbf{Parameters} &  \textbf{Description and Value} \\
 \hline
    Scenario & city in 1000 m*1000 m area \\
 \hline
    VUE velocity& 10, 20, 30, 40, 50 km/h \\
 \hline
    den\_scn & density of SCNs per square kilometer (50 scns/km)\\
 \hline 
   $\theta$ & screening angle of VUE (in degree) \\
 \hline 
   VUE running time & 70000 ms \\
 \hline
    SCN height & 15 meters\\
 \hline
    Communication range & 300 meters\\
 \hline
    Carrier frequency& 6 GHz\\
 \hline
    bw & Bandwidths (10 MHz: 50 PRBs)\\
 \hline
    Transmit power & 30 dB \\
 \hline
    SCN antenna gain & 15 dBi\\
 \hline
    Receiver antenna gain & 0\\
 \hline
    Noise power &  -174 dBm/Hz + $10*\log_{10}$(bw) + 7\\
 \hline
    Pathloss model & pathloss = 128.1 + $37.6*\log_{10}$(Distance) \\
 \hline
    1 tic & 10 ms \\
 \hline
    serving\_scn  & current serving SCN location \\
  \hline
    target\_scn  & target SCN location \\
 \hline
    serving\_sinr & the SINR of VUE from current serving SCN \\
 \hline
    best\_sinr &  the SINR of VUE from the best connection SCN \\
 \hline
    sinr\_min & the minimum SINR value to keep VUE connected to SCN (-7 dB)\\
 \hline 
    avg\_sinr & average SINR of the VUE w.r.t current SCN \\
 \hline
    ho\_avg\_sinr \ & the average SINR value of VUE for each successful handover (dB)\\
 \hline
    best\_cio & cell individual offset (0 dB) \\
 \hline
     current\_cio & cell individual offset (0 dB) \\
 \hline
    ho\_hys & handover hysteresis threshold (3 dB) \\
 \hline
    ho\_timer & handover counter \\
 \hline
    ho\_trigger & handover trigger flag (0 or 1)\\
 \hline
    ho\_exec\_time & handover execution time (25 tics)\\ 
 \hline 
    ho\_times & handover counter for summing the total times of handover \\
 \hline
\end{tabular}
\label{simulation parameters}
\end{center}
\end{table}

\section{simulation results}
A system-level simulator environment is developed using Python to set up the simulation scenario. The simulation parameters can be found in Table \ref{simulation parameters}. The simulation is run for 100 iterations, with the total number of successful handover times being recorded. The density of the SCNs is set at 50 scns/km. The simulation results are used to evaluate the effectiveness of the proposed handover optimization method in reducing the number of handovers and improving the overall network performance.

\begin{figure}[htbp]
	\centering
	\includegraphics[width=\linewidth]{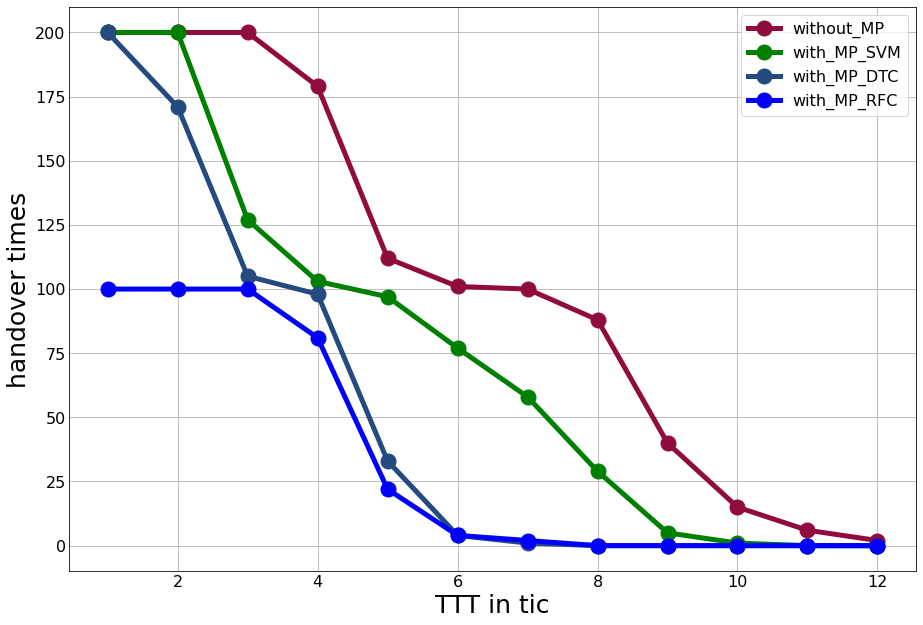}
	\caption{Handover times with 10km/h velocity}
	\label{handover with 10km/h and TTT}
\end{figure}

The results of the simulation are illustrated in Fig. \ref{handover with 10km/h and TTT}, which presents the handover times for different TTT values when using three different ML algorithms (SVM, DTC, and RFC) and a fixed VUE velocity of 10 km/h. Additionally, the handover times for the case without the use of any ML algorithms are provided for comparison.

Starting with a TTT value of 1, the handover times for the cases without MP, with MP using SVM, and with MP using DTC are 200. However, only with MP using RFC, the handover times are reduced to 100 times. As the TTT value increases to 2 tics, the differences in the application of MP with various ML models become more apparent. It can be observed that when the TTT value is relatively small, MP based on SVM and MP based on DTC do not significantly affect the handover times in reducing them. The simulated results in Fig. \ref{handover with 10km/h and TTT} demonstrates that the TTT value plays an important role in the handover reduction. Furthermore, it is evident from the results that as the TTT value increases, the number of handovers reduces significantly, even approaching 0 at 12 tics which means VUE fails in radio connection. This trend can be observed regardless of the application of any ML model, as the VUEs experience a substantial degradation of SINR during the TTT period. This implies that there is an optimal TTT value for a given VUE velocity that can be used to reduce UHO while avoiding handover failure. To compare the effect of MP using ML algorithms and without using ML algorithms on handover performance, when the TTT value is set to 4, the handover times of the traditional method without MP is 179 times. However, when the results of MP based on ML algorithms are considered, the handover times have decreased significantly. For example, when the MP result based on SVM is applied to the simulation, the handover times are reduced to 103, which is almost the same as the handover times obtained using the result of DTC-based MP (112). 

\begin{figure}[htbp]
	\centering
	\includegraphics[width=\linewidth]{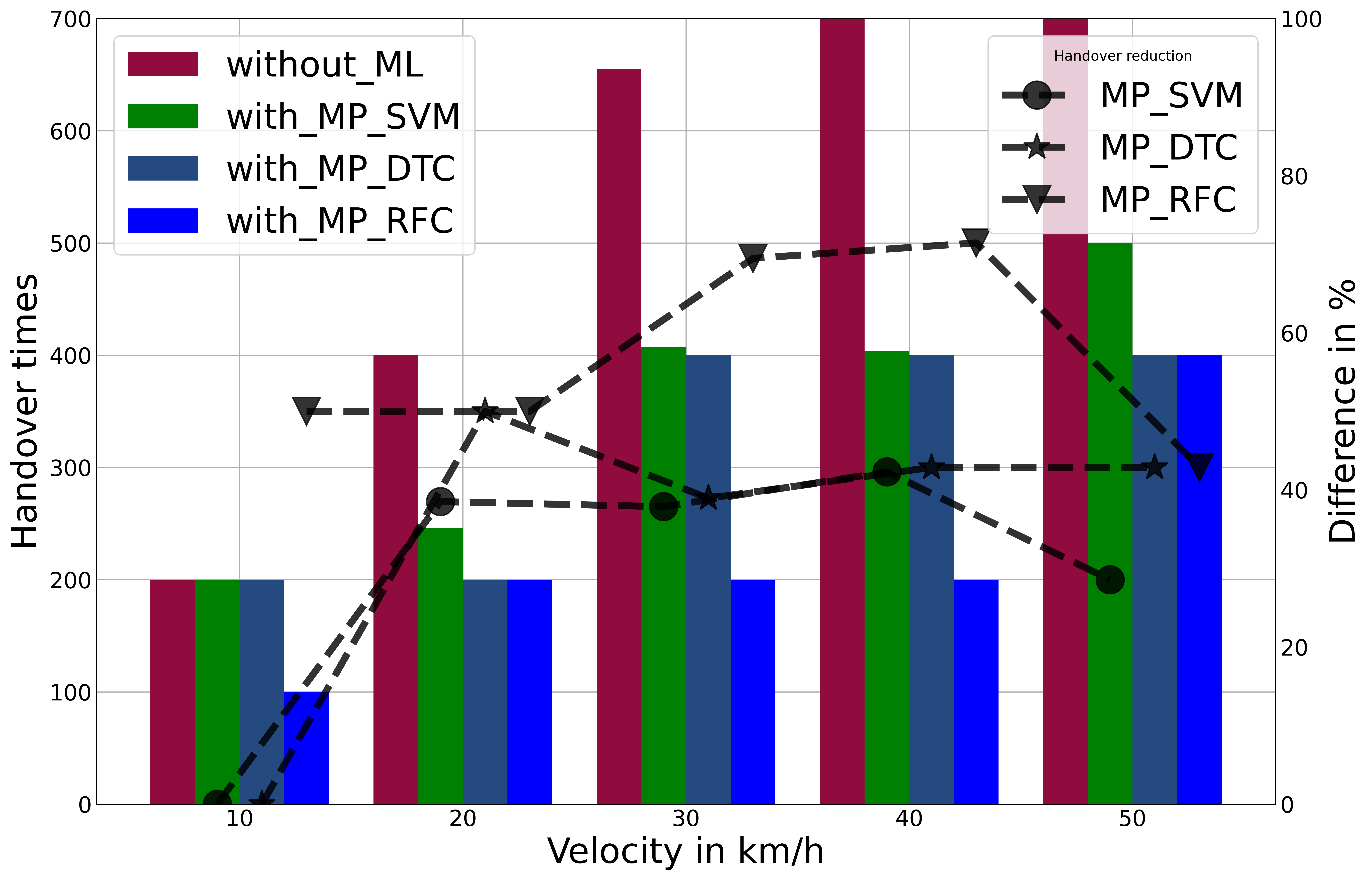}
	\caption{Handover reduction with all velocities}
	\label{handover with all velocities}
\end{figure} 
In Fig. \ref{handover with all velocities}, the effect of varying VUE velocity on handover times is presented. As the velocity increases, the handover procedure becomes more frequent and rapid, making the handover performance more time-critical, especially for real-time services. When no ML algorithms are used in the simulation, and only traditional handover is considered, the handover times increase from 200 times at 10 km/h to 700 times at 50 km/h, as indicated by the red bars. However, when the RFC algorithm is used, the handover times are reduced as compared to the conventional protocol, from 100 times at 10 km/h to 400 times at 50 km/h. This demonstrates that applying ML algorithms to MP can reduce UHO and save time cost.
It is also worth noting that when the vehicle speed is 10 km/h, there is no difference between without ML and with MP using SVM and MP using DTC. This is because, at lower speeds, the handover process is less frequent than in the case of higher speeds, which also shows that MP has a less significant impact on the switching time when the speed of the VUE is relatively low.
The reductions in UHO obtained by applying MP results based on three ML algorithms against traditional handover without any ML are shown as a line chart in Fig. \ref{handover with all velocities}. The average handover reduction ratio for SVM, DTC, and RFC are 29.44\%, 34.93\%, and 56.75\% respectively. When using the RFC algorithm, the greatest decrease in handover times of 71.43\% is obtained at a VUE velocity of 40 km/h. On an average, the use of ML algorithms in MP leads to an reduction of 30.28\% in handover times. Overall, the results of the simulation demonstrate that the proposed handover optimization method using MP with different ML algorithms can significantly reduce the number of handovers and improve the overall network performance.

\section{conclusion}
In summary, the proposed MP method based on ML algorithms has been shown to be effective in reducing UHO in a UDN scenario. By predicting the future route of a VUE, the system can anticipate potential handovers and make more informed decisions to reduce the number of handovers and improve the overall network performance.

Three ML algorithms, SVM, DTC, and RFC, were evaluated and compared for their performance in the simulator. The results show that the RFC algorithm has the highest prediction accuracy and provides the most significant reduction in UHO. In addition, the simulation results also show that the handover performance is affected by various factors such as the TTT and VUE velocity. It is found that larger TTT values may lead to handover failure, but there is a suitable TTT for a given VUE velocity that can also help reduce UHO. Also, the handover times increase with rising velocities, making the handover performance more critical for high-speed VUEs.

In conclusion, the proposed MP method based on ML algorithms can effectively reduce UHO with an average decrease of 30.28\% in handover times and improve the QoS for real-time applications in 5G cellular networks. It is recommended to implement different ML models in different use cases to achieve variant and comparable results. In future work, we plan to implement a more realistic simulation scenario with real data to see the reduction in UHO by MP-based ML algorithms and also to apply the ML algorithms on topics for 6G.

\section{acknowledgement}
This work has been supported by the Federal Ministry of Education and Research of the Federal Republic of Germany (BMBF) as part of the AI4mobile project with funding number 16KIS1170K. The authors would like to appreciate the contributions of all AI4Mobile partners. The authors alone are responsible for the content of the paper which does not necessarily represent the project. 

\bibliographystyle{IEEEtran}
\bibliography{references}

\end{document}